# Near unity Raman β-factor of surface enhanced Raman scattering in a waveguide

Ming Fu[1], Mónica P. dS. P. Mota[1], Xiaofei Xiao[1], Andrea Jacassi[1], Nicholas A. Güsken[1], Yi Li[1,2], Ahad Riaz[1], Stefan A. Maier[1,3], Rupert F. Oulton[1,*]

[1]Blackett Laboratory, Imperial College, Prince Consort Road, London SW7 2BZ, UK
[2]School of Microelectronics, MOE Engineering Research Center of Integrated Circuits for Next Generation Communications, Southern University of Science and Technology, Shenzhen 518055, China
[3]Chair in Hybrid Nanosystems, Nanoinstitute Munich, Faculty of Physics, Ludwig-Maximilians-Universität München, 80539 Munich, Germany
* Corresponding author: r.oulton@imperial.ac.uk

**The Raman scattering of light by molecular vibrations offers a powerful technique to 'fingerprint' molecules via their internal bonds and symmetries. Since Raman scattering is weak[1], methods to enhance, direct and harness it are highly desirable, e.g. through the use of optical cavities[2], waveguides[3–6], and surface enhanced Raman scattering (SERS)[7–9]. While SERS offers dramatic enhancements[6,15,22,2] by localizing light within vanishingly small 'hot-spots' in metallic nanostructures, these tiny interaction volumes are only sensitive to few molecules, yielding weak signals that are difficult to detect[10]. Here, we show that SERS from 4-Aminothiophenol (4-ATP) molecules bonded to a plasmonic gap waveguide is directed into a single mode with $> 99\%$ efficiency. Although sacrificing a confinement dimension, we find $10^4$ SERS enhancement across a broad spectral range enabled by the waveguide's larger sensing volume and non-resonant mode. Remarkably, the waveguide-SERS (W-SERS) is bright enough to image Raman transport across the waveguides exposing the roles of nanofocusing[11–13] and the Purcell effect[14]. Emulating the β-factor from laser physics[15–17], the near unity Raman β-factor observed exposes the SERS technique in a new light and points to alternative routes to controlling Raman scattering. The ability of W-SERS to direct Raman scattering is relevant to Raman sensors based on integrated photonics[7–9] with applications in gas and bio-sensing as well as healthcare.**

Raman spectroscopy has emerged as a powerful technique as it leverages the maturity of laser and detector technologies at visible wavelengths despite its meager efficiency. Various enhancement techniques relying on either stimulated Raman scattering[1] or surface enhanced Raman scattering (SERS)[18–20] have been developed. The stimulated Raman process underpins a range of powerful methods but relies on high intensity and short pulsed optical excitation, which can often damage samples. Meanwhile SERS[21] has become a large field of research exploring metallic nanostructures capable of enhancing Raman by many orders of magnitude, for instance, rough metallic surface[22], nanoparticles[10,23,24], nano-gaps[25,26], waveguides[9,27], and metallic tips[18,28,29].

Despite sensitivity to individual molecules, SERS has several limitations. Firstly, the strongest SERS requires vanishingly small 'hot-spots' where the enhancement is active yet only a few molecules may experience it. Secondly, resonant enhancements restrict the Raman bandwidth. Finally, SERS emerging from localized fields diffracts, making efficient detection difficult[10]. In this letter, we explore waveguide enhanced Raman scattering[3–6], combined with SERS[7–9], using the plasmonic waveguide, shown in **Fig. 1a**. It consists of a plasmonic gap waveguide with optical antenna couplers placed at either end[30–32] on a glass substrate. Raman scattering is enhanced in the gap region by two mechanisms: the increased local excitation intensity by the nanofocusing effect[11–13] and the Purcell effect[14] due to enhanced vacuum fluctuations. Finite difference time domain (FDTD) simulation of the waveguide mode in **Fig. 1b**, shows the optical confinement strength. While the waveguide provides non-resonant SERS over many octaves, the enhancement persists here over the efficiency bandwidth of the antenna-waveguide coupling. Although this approach sacrifices confinement along one direction, strong waveguide-SERS (W-SERS) enables the imaging of Raman transport across a nanostructure and the observation of nanofocusing and Purcell effects. We find that SERS into the gap mode dominates as it drives the Purcell effect. Therefore, we introduce the spontaneous Raman β-factor[15–17], to quantify the proportion of SERS coupling to this single mode. We show that W-SERS produces near unity Raman β-factors with $10^4$ enhancement across a broad spectral range.

In this paper, we explore the Raman scattering of light by 4-ATP molecules[33], which form a self-assembled monolayer on gold-air interfaces, as illustrated in **Fig. 1a**. The optical antennas were designed to provide in and out coupling via a high numerical aperture ($NA = 1.45$) oil immersion microscope system (see **Methods**) over a broad range of wavelengths. FDTD calculations used in the design process are described in **Figure S1**. The microscope system uses a wide-field imaging arrangement, where the focal spot of a pump laser can be scanned across the nanostructure, while the entire field of view is imaged in either Rayleigh or Raman scattering mode. Rayleigh imaging mode evaluates coupling and optical transport of pump light ($\lambda_\text{p} = 790$ nm) across the nanostructure. Meanwhile, Raman imaging mode not only reports Raman spectra ($\lambda_\text{s} = 800 - 900$ nm), but also resolves Raman photon emission and transport across the nanostructure.

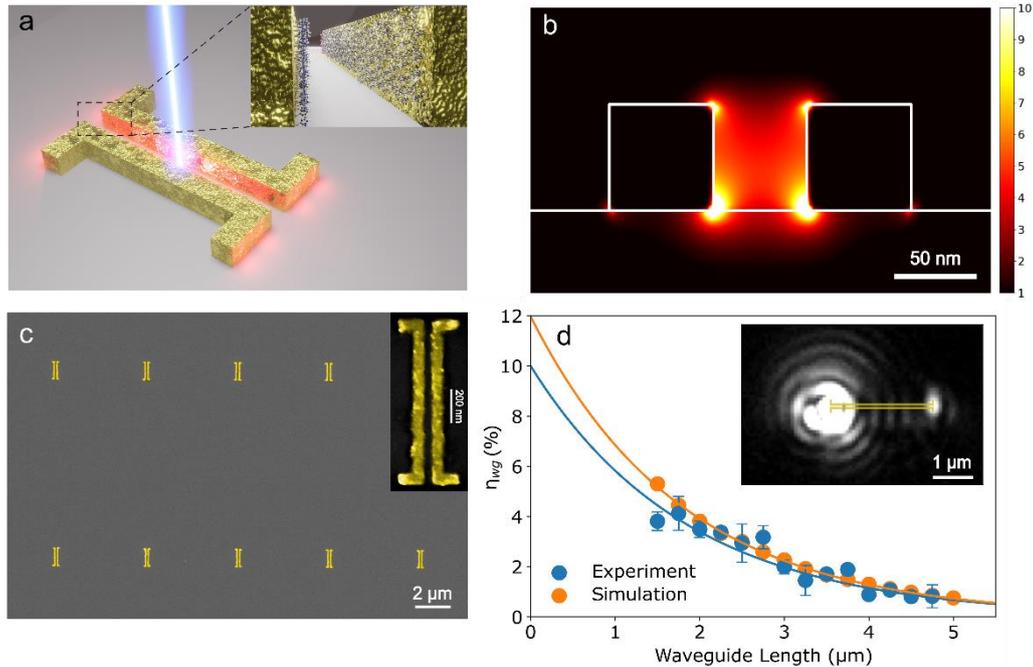

**Fig 1. Illustration of plasmonic gap waveguides and their linear characterisation.** a) Schematic of the plasmonic waveguide bonded with 4-Aminothiophenol (4-ATP) molecules. Under optical excitation of the plasmonic waveguide, most scattered Raman photons can be coupled into the waveguide as surface plasmon polariton (SPP) and coupled out through the antenna pairs at the end of waveguide. While only a few Raman photons can be coupled directly to the free space at the excitation position. b) Simulation of Electricity field intensity $|E|^2$ distribution at the cross section of waveguide. c) Scanning electron microscopic image of the waveguide. The inset panel shows one of the waveguide structures with higher resolution. d) The couple efficiency of waveguide $\eta_{wg}$, which is the ratio between coupled out laser intensity and incident laser intensity, as a function of the waveguide length. The blue dots and orange dots are experiment and simulation results respectively. The solid lines are the fitting results. The insert shows the focused laser excitation image of a waveguide with length of 2.5 µm. The excitation laser was tightly focused on the left side of the waveguide. The yellow line indicates the position of waveguide.

The plasmonic slot waveguides were constructed by two-step Electron Beam Lithography, as described in **Methods**. The high quality and reproducibility of the fabricated waveguides is evident from the Scanning Electron Microscope (SEM) image of an array of devices, shown in **Figure 1c**. From the inset of **Fig. 1c**, the waveguide structure is clearly shown: each plasmonic slot waveguide of length, $L$, and gap width, $G$, is decorated with dimer optical antennas, of length, $l$, on either end. To evaluate the coupling efficiency and propagation loss of these slot waveguides, we undertook a series of linear optical measurements on bare samples without the 4-ATP molecules. Light at a wavelength of $\lambda_\text{p} = 790$ nm was focused to a diffraction limited spot with linear polarization parallel to the antenna to excite one of the antennas. Light coupled into the slot waveguide propagates to the other antenna, where it is imaged in Rayleigh scattering mode. **Figure 1d** shows the total coupling efficiency of waveguide, $\eta_{\text{wg}} = \eta_{\text{antenna}} e^{-L/L_{\text{sp}}}$, of light passing through the plasmonic waveguide for a range of different lengths, $L$, where $\eta_{\text{antenna}} = \eta_{\text{in}} \eta_{\text{out}}$ is the net input and output coupling antenna efficiency, $L_{\text{sp}}$ is the propagation length of the gap plasmon and the gap width, $G = 60$ nm, and antenna length, $l = 150$ nm are fixed. The inset shows an image for $L = 2.5$ µm, where the larger spot is the reflection of the input beam at the input antenna, and the smaller spot is the scattering from the output antenna. Under the assumption that $\eta_{\text{in}}(\lambda_p) \approx \eta_{\text{out}}(\lambda_p)$ we determined both the antenna-waveguide coupling efficiency, $\eta_{\text{in}}(\lambda_p) \approx \sqrt{\eta_{\text{antenna}}} = 30 \pm 2$ %, and $L_{\text{sp}}(\lambda_p) = 2.1 \pm 0.2 \ \mu m$. At Raman wavelengths, we find $\eta_{\text{out}}(\lambda_s = 860) = 0.46 \pm 0.01$ (**Figure S2**) with sustained out-coupling efficiency achievable over the Raman emission band, as shown by simulations in **Figure S1**.

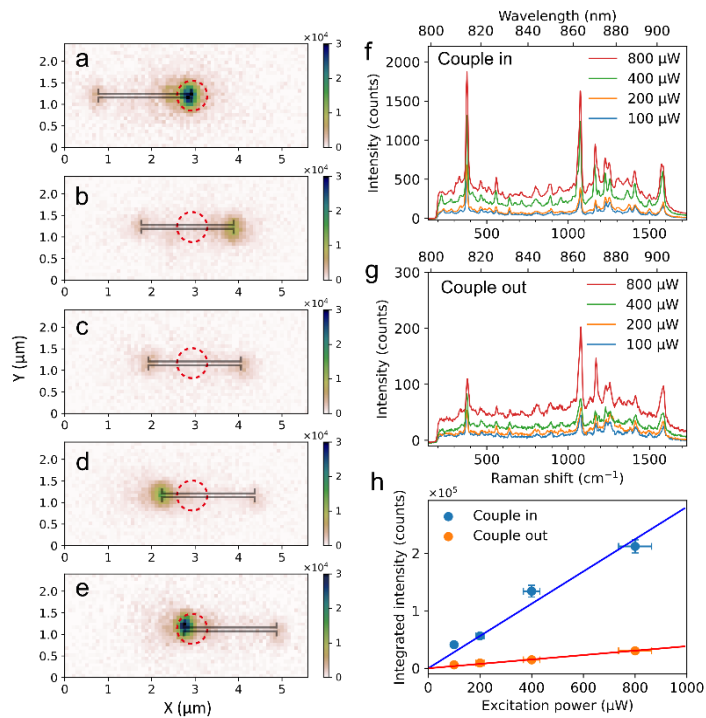

**Fig 2**. **Imaging and spectroscopy of waveguide-based surface enhanced Raman scattering**. a-e) Raman image of molecules bonded on the surface of plasmonic waveguide, by tightly focusing the laser on different position of the waveguide (L= 2 µm). The dashed red circle indicates the position of the excitation laser spot. To guide the eye, the grey solid lines indicate the position of waveguide. f) and g) are the Raman spectra extracted from the left side (laser couple in) and right side (laser couple out) of waveguide with L = 2 µm, with different excitation powers. h) Integrated Raman scattering intensity of laser couple in and couple out positions as a function of excitation power. The solid lines are the linear fitting results.

To exploit the enhancement and control of Raman scattering by the plasmonic waveguide, 4-ATP molecules were chemically bonded to the surface of the Au nanostructure. The details of the sample preparation are described in **Methods**. **Figure 2** shows results of our experiment in Raman imaging mode for a pump wavelength of $\lambda = 790$ nm. The series of images in **Figs. 2a-e** show the result of scanning the focal spot across the nanostructure from one antenna to the other. The brightest Raman emission occurs from an input antenna when the pump coincides with that antenna (**Figs 2a and 2e**). The spectra of Raman scattering from both sides of the waveguide were measured by using a sensitive spectrometer, with spectra shown in **Figs. 2f** and **2g**. The broadband Raman scattering with featured Raman peaks of 4-ATP can be clearly observed from the spectra of both the input and output coupler positions. The integrated intensities of Raman scattering from both sides of the waveguide were found to be linearly dependent on the power of the excitation laser (**Figure 2h**). In addition, we also measured 4-ATP molecules chemically bonded to a flat Au film, where a Raman spectrum was barely visible under the same experimental conditions. With a larger pump power and longer integration time, we were able to estimate that the plasmonic waveguide provided $10^4$ times more Raman scattering compared to the control sample (**Figure S3**).

The waveguide was then shifted from left to right, while the position of the focused laser beam was fixed, indicated by the dashed circle in the images of **Figs. 2a-e**. Raman scattering on the right-hand side gradually decreased when the waveguide was moved to the right, until eventually equal Raman emission was observed from both antennas when the laser beam was near the center of the waveguide (**Figure 2c**). With further displacement of the waveguide to the right, the reverse of the image in **Figure 2a** was observed, where the excitation beam is now on the left side of the waveguide (**Figure 2e**). The reversable position dependent Raman scattering shows the high precision of the home-built microscope and further confirms the high quality of the fabricated waveguides.

There are two key observations to draw from this experiment. Firstly, the Raman scattering is clearly emanating from regions not directly excited by the pump laser in all images in **Figure 2**. In **Figs. 2a** and **2e**, transport is observed over a distance of 2 μm from the diffraction limited excitation spot, which has $e^{-2}$ diameter of $0.82\lambda_\mathrm{p}/NA \approx 447$ nm. Secondly, very little Raman scattering emanates from the position of the excitation beam in **Figure 2c** when the waveguide is centrally pumped, Raman scattering is only observed at the output couplers. This is evidence of the strengthened vacuum fluctuations of the slot waveguide, where Raman scattering into the waveguide is accelerated compared to Raman scattering into radiation modes of free space. This causes Raman scattering into the waveguide modes to dominate over that scattered as radiation leading to a high Raman β-factor. The broadband nature of this enhancement mechanism explains why no Raman, or indeed gold fluorescence, is observed at the position of the excitation beam.

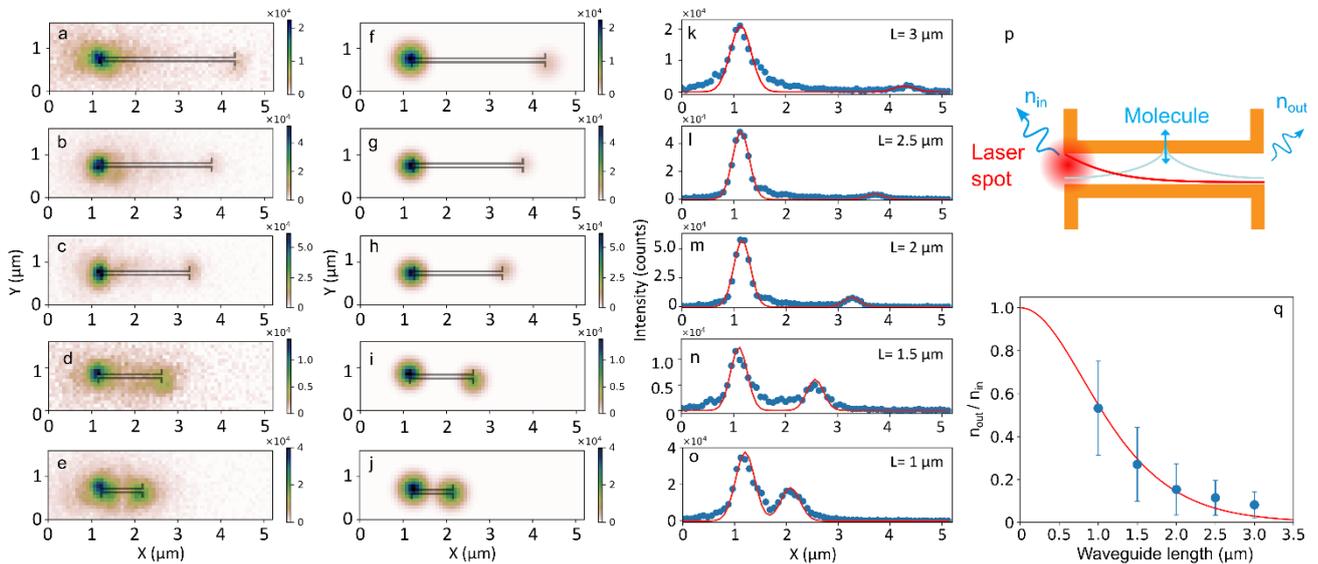

**Fig 3. Imaging of Raman scattering transport in W-SERS**. a-e) The Raman image of molecules bonded to waveguide with different lengths, with the excitation beam tightly focused on the left side of the waveguide. f-j) The two-dimensional Gaussian fittings of the a-e). To guide the eye, the grey solid lines indicate the position of waveguide. k-o) the image profile of a-e) and f-j). p) The scheme of laser propagation along the waveguide and Raman scattering of molecules. $n_\mathrm{in}$ and $n_\mathrm{out}$ are the Raman photons collected from laser couple in and couple out positions. q) The ratio of Raman photons collected from laser couple in and couple out positions $n_\mathrm{out}/n_\mathrm{in}$, with laser illuminated at the antennas. The red solid line is the theoretical result.

To explore the mechanisms of Raman scattering in the plasmonic gap waveguides, the excitation laser was positioned on the left-hand antenna of waveguides of varying length, $L$, and Raman scattering images were collected. As observed in **Figs 3 a-e**, the Raman scattering is strongest at the position of the excitation laser, while the Raman scattering on the right-hand side decreases with increasing waveguide length. To quantitively determine the intensity of Raman scattering on both sides of the waveguide, the Raman images were fit to 2D Gaussian functions, as shown in **Figs 3 f-j**, where the matching profiles of raw and fitted images confirm the quality of the fits (**Figs 3 k-o**). We have developed a theoretical model that considers the bi-directional Raman scattering into the waveguide mode as shown in the scheme of **Figure 3p** (see **Methods**). For excitation at the left-hand antenna at position $x = 0$, for a waveguide of length $L$, the ratio of the Raman photon numbers from the input, $n_{\text{in}}$, and the output, $n_{\text{out}}$, antennas is,

$$\frac{n_{\text{out}}}{n_{\text{in}}} = \bar{L}[\sinh \bar{L}]^{-1}, \tag{1}$$

where $\bar{L} = L/L_{\text{sp}}$, is the ratio of waveguide and mode propagation lengths and is the only free parameter. The model matches well with the experimental data (**Figure 3q** and **Figure S5**) for a propagation length of $0.7 \pm 0.1\ \mu m$, which is about $2.6 \times$ smaller than the value obtained from Rayleigh imaging measurements, in **Fig 1d**. To confirm this modified value of $L_{sp}$, extracted from the Raman model, we directly measured $L_{sp}$ again for the samples coated with 4-ATP using Rayleigh scattering imaging mode. We find that the 4-ATP introduces additional absorption, reducing propagation length, but not affecting the coupling efficiency (**Figure S2**). The role of molecular absorption on Raman scattering is clearly shown in this waveguide setting and indicates the length scales over which Raman scattering can be effectively collected. In this case, the total Raman signal is maximized for increasing slot waveguide length up to about $L = 2L_{sp} \approx 1.4\ \mu m$ (see Section S4 in Supplementary Information), which presents a substantial increase in interaction volume compared to other SERS approaches. For molecules that absorb less pump light, greater interaction lengths would be possible.

Alternatively, when the waveguide is illuminated centrally, we consider the total number of Raman photons collected from both antennas, $n^{(c)}$. When normalized against $n_{\text{in}}$, the number of photons emitted from the input antenna for the antenna-coupled case, we find a ratio

$$\frac{n_{\text{in}}}{n^{(c)}} = \left(\frac{A_0}{A_{\text{wg}}(\lambda_{\text{p}})}\right) \frac{\eta_{\text{in}}(\lambda_{\text{p}}) n_{\text{g}}(\lambda_{\text{p}}) L_{\text{sp}}}{4\sqrt{2\pi}\rho H} e^{\bar{L}/2}, \tag{2}$$

where $\rho = 0.41\lambda_{\text{p}}/NA$ is the radius of the Gaussian beam used to excite the waveguide, both centrally and at the antenna, $A_0 = \pi\rho^2$ is the beam area, and $H$ is the local electric field enhancement of the Gaussian beam due to the waveguide's narrow gap. $H$ was evaluated numerically as described in **Figure S4**. The experimentally measured ratio is fit to the function, $n_{\text{in}}/n^{(c)} = ae^{L/2L_{\text{sp}}}$, for varying waveguide length, $L$, as shown in **Figure S6**. The value $a = 0.97 \pm 0.01$ allows us to determine experimentally the plasmonic waveguide mode area, and the average Stoke's field propagation length, $L_{\text{sp}}(\lambda_{\text{s}}) = 0.780 \pm 0.004\ \mu\text{m}$. The mode area reduction, $A_0/A_{\text{wg}}(\lambda_{\text{p}}) = 14.94 \pm 0.96$, compares well with the simulated value of 14.68 (**Table S1**). Notably, the Raman signal from antenna excitation always dominates over central excitation.

We can use this estimate of the waveguide mode area to determine the average pump intensity enhancement, $I_{\text{p0}}/I_{\text{f0}} = 84.45 \pm 0.87$, relative to the peak Gaussian beam intensity, $I_{\text{f0}}$, of the gold film control sample. When multiplied with a numerical estimate for the Raman Purcell enhancement factor of $F_{\text{p}} = 42.55$, we find the total Raman enhancement factor of $7{,}504 \pm 78$, which compares well to the observed enhancement relative to the gold film control sample of $13{,}032 \pm 227$ (**Figure S3**).

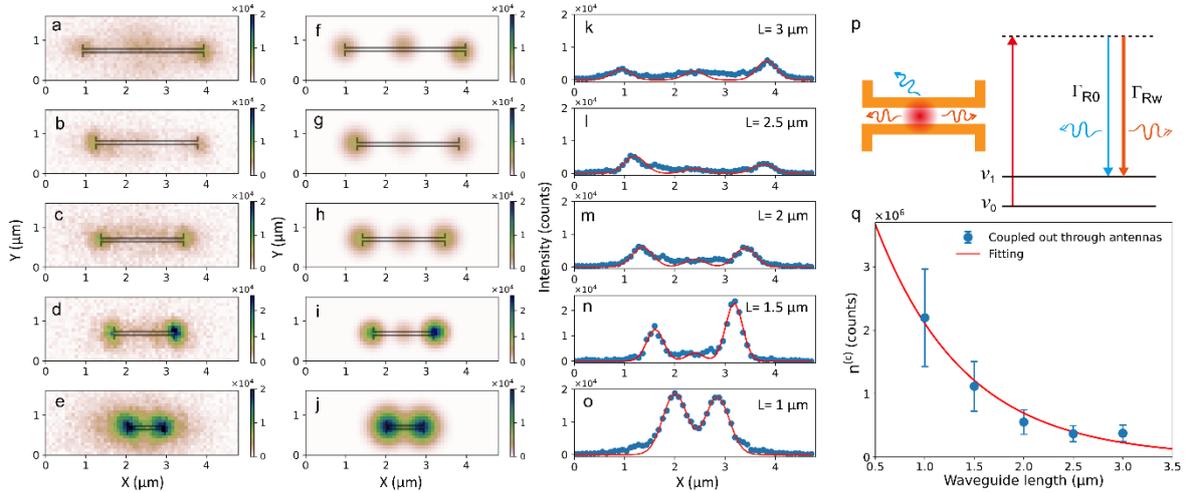

**Fig 4. Experimental evaluation of the Raman $\beta$ factor**. a-e) The Raman image of molecules bonded to waveguide with different lengths, with the excitation beam tightly focused on the middle of the waveguide. f-j) The two-dimensional Gaussian fittings of the a-e). The grey solid lines indicate the position of waveguide. k-o) the image profile of a-e) and f-j). p) The scheme of Raman scattering coupled to free space and to the waveguide as surface plasmon polariton, with scattering rate of $\Gamma_{R0}$ and $\Gamma_{Rw}$. q) The intensity of Raman scattering coupled out through two antenna pairs $n^{(c)}$ as a function the waveguide length. The red solid line the exponential fitting.

We now continue with the centrally illuminated waveguide case and consider the proportions of Raman photons emitted directly as radiation and emitted from the antennas via the waveguide mode. Let us assume that the 4-ATP molecules excited by the pump beam scatter Raman photons directly as radiation at a rate, $\Gamma_{R0}$, and into the waveguide mode with rate $\Gamma_{Rw}$, as illustrated in the schematic of **Figure 4p**. Using our Raman imaging approach, we can now estimate the Raman β-factor, where $\beta \approx \Gamma_{Rw}/(\Gamma_{Rw} + \Gamma_{R0})$. To estimate to Raman β-factor, we measured the Raman photons directly coupled to free space at the excitation position, $n^{(c)}_{R0} \propto \Gamma_{R0}$, and compare this with the total Raman photons collected at the antennas, $n^{(c)}$. The number of Raman photons emitted directly into the waveguide mode at the excitation position, $n^{(c)}_{Rw} \propto \Gamma_{Rw}$, are related to $n^{(c)}$ through the propagation loss to the antennas and their out coupling efficiency, where $n^{(c)} = \eta_{out}(\lambda_s) n^{(c)}_{Rw} e^{-\bar{L}/2}$. We thus estimate the Raman β-factor using,

$$\beta = \left[1 + \frac{n^{(c)}_{R0}}{n^{(c)}_{Rw}}\right]^{-1} \qquad (3)$$

From the fitting, shown in **Figure 4q**, we found $\beta = 0.993 \pm 0.001$ so that with a high degree of certainty, less than 1% of the Raman emitted at the excitation spot radiate directly. We can check this value of β against our calculated Purcell factor of $F_p = 42.55$, with $\beta \approx (F_p - 1)/F_p = 0.98$ (see **Methods**). Raman scattering is therefore clearly bidirectional emission with near unity efficiency and couples to a single waveguide mode. We note that a single mode silica optical fiber would have a Raman $\beta \sim 10^{-3}$. As is seen in low threshold lasers, the Raman β factor determines the strength of stimulated Raman scattering. We thus propose that the W-SERS approach reduces the threshold of stimulated Raman scattering by many orders of magnitude.

W-SERS enables strong Raman enhancement due to lateral confinement, with the added ability to efficiently collect the scattering in a waveguide with near unity Raman β-factor. Although we observed $10^4$ times enhancement, the enhancement scales dramatically with reducing gap width so that $> 10^6$ is possible for a gap width of 10 nm[25] (**Figure S7**). Here we have seen Raman scattering from the relatively sparse distribution of 4-ATP molecules bound to the gold surface; however, the gap region presents a substantial sensing volume with the opportunity for stronger optical-molecule interactions. The W-SERS technique also provides new insight into the SERS mechanism and in particular the role of the Raman β-factor. We have proposed that this could reduce the threshold for stimulated SERS by many orders of magnitude, a route to improved Raman sensing. The ability for W-SERS to direct Raman scattering with a high Raman β-factor could be exploited for highly sensitive on-chip Raman sensor based on integrated photonics with applications in gas and bio-sensing as well as healthcare applications.

**Methods:**

**Sample preparation**

**Fabrication of Antenna/Waveguide**
The plasmonic waveguides were fabricated by electron beam lithography (EBL) on 175 μm thick glass substrates. The substrate was submerged for 24 hours in a solution of DECON and water (1:100), and then dipped in acetone, sonicated, rinsed with IPA, dried with nitrogen and plasma ashed with oxygen for 5 min. A polymer, resist PMMA 475, was spin-coated on the glass substrate (3500 rpm, 1 min) and baked at 180 C for 5 min. On top of the first polymer layer a second one of PMMA 950 is spin-coated and then baked with the same parameters of the first one. To guarantee a high conductivity of the sample, a layer of E-Spacer was spin-coated as well (2000 rpm, 1 min) and baked (90 ℃, 1 min). A Raith Nanofabrication Electron Micoscope was used to perform the electron lithography. A lift-off process was performed to remove the polymer and the gold on top of it. The sample was sunk into acetone for 24 hours then rinsed with IPA and dried with nitrogen. To create the most reproducible gaps, a two-step method was employed for each half of the waveguide design using the same steps described above. For alignment of the two EBL exposures, markers were created during the first step and an automatic alignment procedure guaranteed excellent matching of the structures.

**Coating 4-ATP molecules on gold waveguides**
Before coating 4-ATP molecules on gold waveguides, the waveguides sample was rinsed with Acetone and IPA. Then the sample was dried out with $N_2$ flow. 4-ATP solution was prepared by dispersing the 4-ATP molecules in ethanol. The waveguides sample was further cleaned in $O_2$ plasma for more than 1 minute. Then the sample was immediately dropped into the prepared 4-ATP ethanol solution and kept in the solution overnight. Lastly, the sample was rinsed with ethanol before the measurements.

**Optical Characterization of Waveguides**

The individual waveguide structures with and without 4-ATP molecules were characterised by a home-made wide-field optical microscope system. The coherent light from a tuneable continuous-wave Ti-Sapphire laser system (Model 3900, Spectra-Physics) was tightly focused onto the sample with diffraction limited spot size by an oil objective lens (Plan Apo Lambda, 100 ×, NA=1.45, Nikon). A half waveplate was used to adjust the beam's polarisation to be parallel to the antennas of the waveguides. A short-pass filter (FESH0800, Thorlabs) was used to clean the excitation laser frequency. For laser scattering measurement, a wedge beam splitter was used to reflect the laser beam to the objective lens whilst allowing the laser scattering signal from the sample transmits through it with high efficiency. For Raman image and spectra measurements, a long pass dichroic beam splitter (FF801-Di02-25x36, Semrock) was used to reflect the laser beam while the Raman scattering signal can transmit through efficiently. The sample was mounted on a 3-dimensional open loop piezo stage (3-Axis NanoMax, Thorlabs) with stepper motor (MAX383, Thorlabs). One to two drops of immersion oil with refractive index n = 1.45 (Nikon type N) were placed in between the substrate and the objective lens to minimise the refractive index mismatch.

For navigation purposes, a white light source (QTH10, Thorlabs) illuminated the side of the waveguide structures, so that their images and the transmitted white light could be collected on the other side by a camara (webcam, Logitech). A long pass filter (BLP01-808R-25, Semrock) and a notch filter (NF785-33, Thorlabs) were placed to block the laser scattering for Raman image and spectra measurement. An electron multiplying charge coupled device (EMCCD) (Rolera EM-C$^2$) was placed on the image plane to image the laser excitation image and Raman image. The Raman photons are sent to a spectrometer (Acton 300, Princeton Instrument) combined with a cooled CCD camera (PIXIS 100, Princeton Instrument) to measure the Raman spectra.

**Waveguide Enhanced Raman Theory**

We consider Raman scattering in a waveguide using a scalar 1D differential equation in the number of scattered Raman photons[34], described in detail in the **Supplementary Information**. The generated Stoke's Raman photon numbers along the waveguide, $n_f(z)$ and $n_b(z)$, in the forwards and backwards directions are determined by the 1D differential equations,

$$\frac{dn_f(z)}{dz} = -\alpha_s n_f(z) + g_{Rw} I_p(z) \tag{4}$$

$$\frac{dn_b(z)}{dz} = \alpha_s n_b(z) - g_{Rw} I_p(z) \tag{5}$$

where $g_{Rw}$ is the forward and backward Raman scattering coefficient into the waveguide mode, and $\alpha_s = L_{sp}^{-1}$, is the absorption coefficient at the Stokes wavelength. Pump light is introduced either via waveguide coupling from the antenna, such that,

$$I_p(z) = I_{p0} e^{-\bar{L}} \tag{6}$$

where $I_{p0}$ is the local peak pump intensity in the waveguide mode at the input coupler and $\bar{L} = \alpha_s L = L/L_{sp}$. In this case, the numbers of Raman photons emitted from each antenna coupler are,

$$n_{out} = \eta_{out}(\lambda_s) n_f(L) = \eta_{out} \kappa L e^{-\bar{L}} \tag{7}$$

$$n_{in} = \eta_{out}(\lambda_s) n_b(0) = \eta_{out}(\lambda_s) \frac{\kappa}{2\alpha_s}\left(1 - e^{-2\bar{L}}\right) \tag{8}$$

Where $\eta_{out}$ is the output coupling efficiency from the waveguide antennas at the Stoke's Raman wavelength and $\kappa = g_{Rw} I_{p0}$. The ratio of Raman scattering from the two couplers is thus,

$$\frac{n_{out}}{n_{in}} = \bar{L}[\sinh \bar{L}]^{-1} \tag{9}$$

When the waveguide is illuminated centrally, we consider the following Gaussian intensity distribution,

$$I_p(z) = H I_{g0} e^{-2(z-L/2)^2/\rho^2} \tag{10}$$

where $\rho$ is the Gaussian beam radius and $I_{g0}$ is the peak intensity. The parameter, $H = 4.29$, accounts for the local electric field enhancement of the Gaussian beam due to the waveguide's narrow gap and was evaluated numerically as described in the **Supplementary Information**. The total number of Raman photons collected from both antennas is,

$$n^{(c)} = \eta_{out}(\lambda_s)\left(n_f^{(c)}(L) + n_b^{(c)}(0)\right) = \eta_{out}(\lambda_s)\sqrt{2\pi}\rho\gamma H e^{\alpha_s^2 \rho^2/8 - \bar{L}/2} \tag{11}$$

where $\gamma = g_{Rw} I_{g0}$. When comparing Raman scattered photons by the antenna and centre illumination methods, we account for the same incident power, $P_0$. For the waveguide mode, the coupling peak intensity to the waveguide is $I_{p0} = \eta_{in}(\lambda_p) n_g(\lambda_p) P_0 / A_{wg}(\lambda_p)$, where $n_g(\lambda_p)$ is the mode group index and $A_{wg}(\lambda_p)$ is the waveguide mode area at the pump wavelength. For the Gaussian beam intensity, $I_{g0} = 2P_0/A_0$, where $A_0 = \pi \rho^2$. The ratio $n_{\text{in}}/n^{(c)}$, which can be experimentally measured, allows us to determine the ratio of the Gaussian beam and waveguide mode areas:

$$\frac{A_0}{A_{\text{wg}}(\lambda_{\text{p}})} = \left(\frac{n_{\text{in}}}{n^{(c)}}\right) 4\sqrt{2\pi} \alpha_s \rho \frac{H}{\eta_{\text{in}}(\lambda_{\text{p}}) n_{\text{g}}(\lambda_{\text{p}})} e^{-\bar{L}/2} \quad (12)$$

which agrees well with computed mode areas as shown in the **Table S2**. The Purcell factor at the Stoke's wavelength can be computed using the formula[35–37],

$$F_{\text{pw}} = 3 \frac{n_{\text{g}}(\lambda_s)}{\pi} \frac{(\lambda/2)^2}{A_{wg}(\lambda_s)} \quad (13)$$

where $n_{\text{g}}(\lambda_s)$, the mode group index, and $A_{\text{wg}}(\lambda_s)$ must be computed numerically at the Stoke's wavelength. The total Purcell factor must also include the Raman scattering to radiation. Here we assume that Raman scattering to radiation on balance is not affected by the waveguide so that $F_p \approx 1 + F_{\text{pw}}$. The estimation of the total Raman enhancement factor is discussed in the **Supplementary Information**. The fraction of Raman scattering directly into the waveguide mode (Raman $\beta$ factor), may be crudely approximated by,

$$\beta \approx \frac{F_{\text{pw}}}{1 + F_{\text{pw}}} \approx \frac{F_p - 1}{F_p} \quad (14)$$

which is often quoted in the literature in the case of spontaneous fluorescence[15].

## Funding

This work was supported by the EPSRC Reactive Plasmonics Programme (EP/M013812/1) and the Leverhulme Trust. This project has received funding from the European Union's Horizon 2020 research and innovation programme under the Marie Skłodowska-Curie grant agreement No 844591.

## Author Contributions

M.F., M.P.dS.P.M. and R.F.O. developed the idea and designed the experiments. M.P.dS.P.M. and X.X. designed and simulated the waveguide structures and their mode properties. A.J., N.A.G. and Y.L. fabricated the waveguides and performed structural characterisation. M.F., M.P.dS.P.M. and Y.L. prepared for the sample with 4-ATP molecules. M.F. and M.P.dS.P.M. conducted the experiments and analysed the data. A.R. and R.F.O. developed the waveguide Raman model and analysed this in the context of the experimental data. M.F., M.P.dS.P.M., X.X., A.J., N.A.G., Y.L., A.R., S.A.M. and R.F.O. all contributed to the writing of the manuscript.

**Supplementary Information for "Near unity Raman $\beta$-factor of surface enhanced Raman scattering in a waveguide"**

Ming Fu[1], Mónica P. dS. P. Mota[1], Xiaofei Xiao[1], Andrea Jacassi[1], Nicholas A. Güsken[1], Yi Li[1,2], Ahad Riaz[1], Stefan A. Maier[1,3], Rupert F. Oulton[1,*]

[1]*Blackett Laboratory, Imperial College, Prince Consort Road, London SW7 2BZ, UK*
[2]*School of Microelectronics, MOE Engineering Research Center of Integrated Circuits for Next Generation Communications, Southern University of Science and Technology, Shenzhen 518055, China*
[3]*Chair in Hybrid Nanosystems, Nanoinstitute Munich, Faculty of Physics, Ludwig-Maximilians-Universität München, 80539 Munich, Germany*
\* Corresponding author: r.oulton@imperial.ac.uk

## Table of contents

## 1. Simulation of Plasmonic Gap Waveguide and Coupler

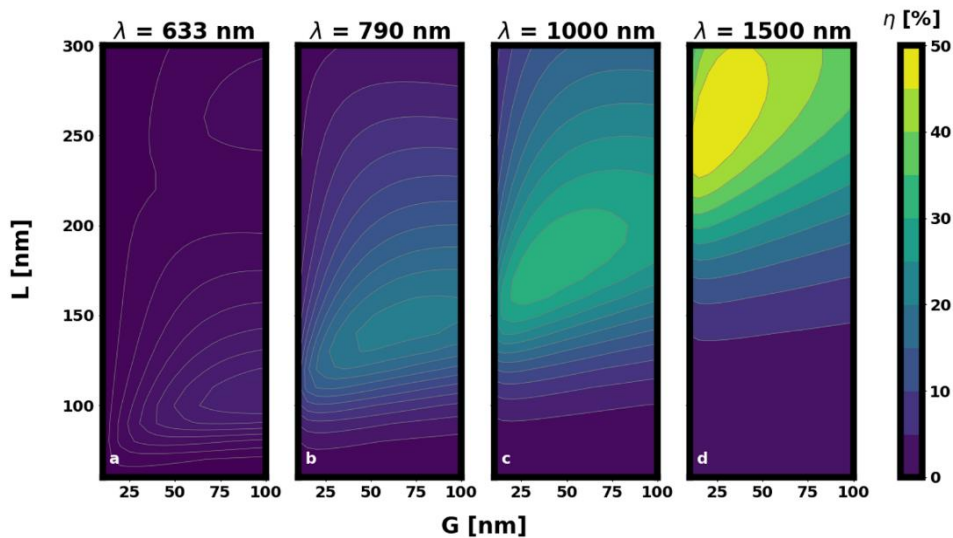

**Figure S1.** Simulated maximum incoupling efficiency $\eta_{in}$ in for the "infinite" perfectly matched layers (PML) system for a range of antenna lengths and gap widths at (a) $\lambda$ = 633 nm, (b) 790 nm, (c) 1000 nm and (d) 1500 nm.

## 2. Propagation Length of Gap Waveguides Coated with 4-ATP Molecules

Raman scattering of the gap waveguides coated with 4-ATP was found to be reduced compared to waveguide without molecules. Thus, further Rayleigh scattering measurements were undertaken to confirm molecular absorption as the origin of the discrepancy. Indeed, Figure S3 confirms that the propagation length in these waveguides is affected by the 4-ATP at both the pump and Raman wavelengths. Attenuation is found to be greater at the pump wavelength and this is consistent with experimental findings.

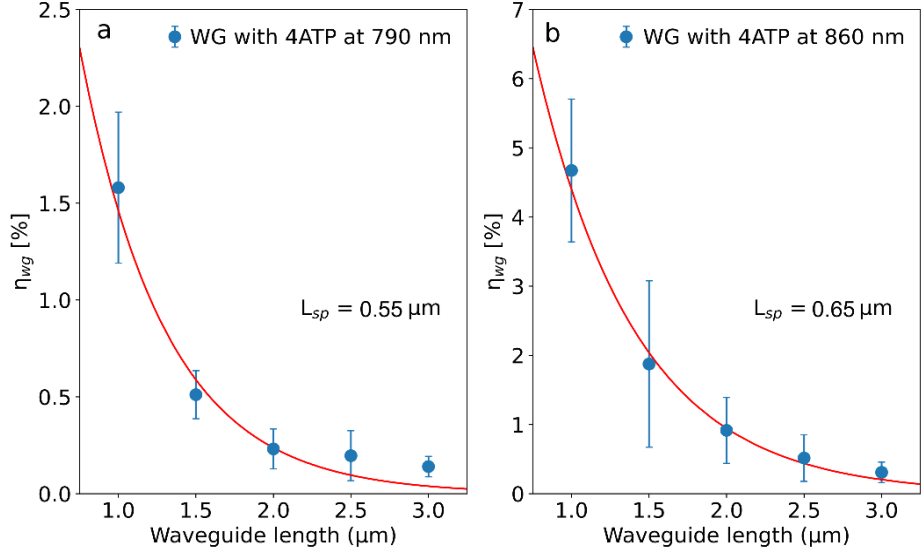

**Figure S2.** The couple efficiency of waveguide $\eta_{wg}$ coated with 4-ATP molecules, as a function of the waveguide length. The surface plasmon propagation lengths are $0.55 \pm 0.04\ \mu m$ and $0.65 \pm 0.05\ \mu m$ respectively, with the incident laser wavelength of (a) 790 nm and (b) 860 nm.

From the fitting of Figure S3, we find the antenna coupling efficiency at 790 nm $\eta_{in}(\lambda = 790\ nm) = \eta_{out}(790\ nm) = 0.31 \pm 0.01$, which is consistent with the waveguide without 4-ATP molecules. The antenna coupling efficiency at 860 nm is $\eta_{in}(\lambda = 860\ nm) = \eta_{out}(860\ nm) = 0.46 \pm 0.01$. Note that the extracted propagation lengths here are lower than observed in Raman scattering measurements.

## 3. Raman Scattering of 4-ATP Molecules Coated on Au Film

Raman scattering exited and collected from an input antenna of a waveguide was directly compared with Raman scattering from a planar gold film. Data are shown in Figure S2. Signal from the gold film is significantly weaker despite similar numbers of molecules over the beam area being excited. The beam is focused though an air objective (60 ×, NA=0.95) onto the gold film, $\rho = 0.41\lambda/NA = 0.34\ \mu m$, giving a focal spot are, $\pi\rho^2 \approx 0.36\ \mu m^2$. Meanwhile the active area is taken for an optimal waveguide length for Raman scattering, $L = 2L_{sp} = 1.4\ \mu m$, as described in Section S4. The gap waveguide active area is thus approximately twice the gap length times gap height, $2 \times 2L_{sp} \times 0.06 = 0.17\ \mu m^2$. The value of $L_{sp} = 0.7 \pm 0.1\ \mu m$ was taken from the fitting in Fig. 3q. The intensity ratio of the two techniques from Figure S2 of $6147 \pm 107$, is normalised by the ratio of active areas, $R_{\text{mol}} \approx 0.36/0.17 = 2.12$, to give the enhancement of $13{,}032 \pm 227$.

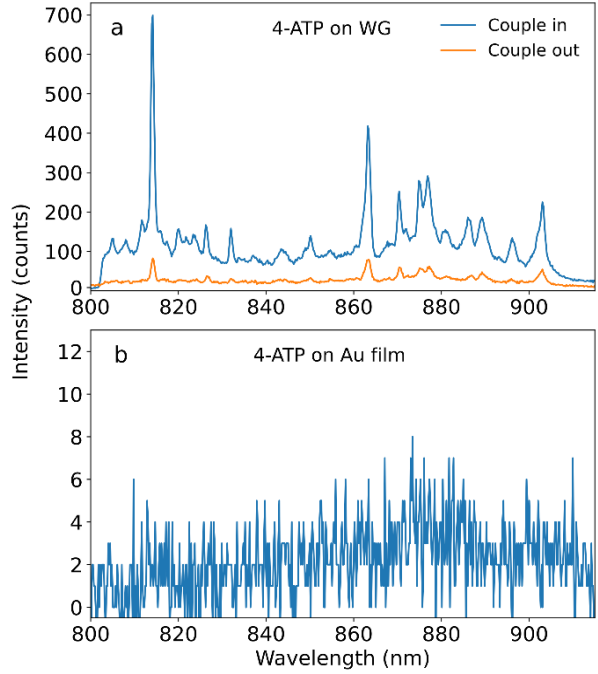

**Figure S3.** (a) Raman spectra of 4-ATP molecules coated on plasmonic waveguide with length $= 2\ \mu m$, excited by 790 nm laser with pump power of $200\ \mu W$, integration time of 10 s. (b) Raman spectrum of 4-ATP coated on Au film, excited by 790 nm laser with power of 7.5 mW, integration time of 30 s. The Raman scattering intensity ratio of 4-ATP coated on waveguide over Au film is around 6147.

## 4. Theory of Raman Scattering in a Plasmonic Waveguide

We consider Raman scattering in a waveguide as a scalar 1D differential equation in the number of scattered Raman photons. The coupling efficiency into and out of the waveguide is assumed to be $\eta_{\text{in}}$ and $\eta_{\text{out}}$, and is considered separately. The generated Stokes Raman photon numbers along the waveguide, $n_{\text{f}}(z)$ and $n_{\text{b}}(z)$, in the forwards and backwards directions are determined by the 1D differential equations,

$$\frac{dn_{\text{f}}(z)}{dz} = -\alpha_{\text{s}} n_{\text{f}}(z) + g_{\text{Rw}} I_{\text{p}}(z),$$

$$\frac{dn_{\text{b}}(z)}{dz} = \alpha_{\text{s}} n_{\text{b}}(z) - g_{\text{Rw}} I_{\text{p}}(z),$$

where $g_{\text{Rw}}$ is the rate of forward and backward Raman scattering into the waveguide mode, and $\alpha_{\text{s}} = L_{\text{sp}}^{-1}$, is the absorption coefficient at the Stokes wavelength, and where $L_{\text{sp}}$ is the propagation distance. Pump light is introduced as a 1D distribution, $I_{\text{p}}(z)$, to be determined. Here, we will consider two illumination conditions: direct excitation by coupling pump light to the waveguide mode via an antenna; and indirect excitation by illuminating the centre of the waveguide with a focussed Gaussian beam. Both cases are described in the main text.

**Direct optical pumping via waveguide coupling**

The optical pump beam can be directly coupled to the same waveguide mode as the Raman scattering, by illuminating the antenna couplers, as described in the main text. The pump intensity distribution can be determined using the following 1D differential equation:

$$\frac{dI_\mathrm{p}(z)}{dz} = -\alpha_\mathrm{p} I_\mathrm{p}(z) - g_\mathrm{R} I_\mathrm{p}(z) \approx -\alpha_\mathrm{p} I_\mathrm{p}(z),$$

where $\alpha_\mathrm{p}$ is then pump absorption coefficient in the waveguide. Here we assume that the total Raman scattering coefficient, $g_\mathrm{R}$, is weak and so loss is dominated by waveguide absorption. The optical pump distribution in this case is:

$$I_\mathrm{p}(z) = I_\mathrm{p0} e^{-\alpha_\mathrm{p} z},$$

where $I_\mathrm{p0}$ is the local peak pump intensity in the waveguide mode at the input coupler. The differential equations for Stoke's scattered photons become:

$$\frac{dn_\mathrm{f}(z)}{dz} + \alpha_\mathrm{s} n_\mathrm{f}(z) = \kappa e^{-\alpha_\mathrm{p} z},$$

$$\frac{dn_\mathrm{b}(z)}{dz} - \alpha_\mathrm{s} n_\mathrm{b}(z) = -\kappa e^{-\alpha_\mathrm{p} z},$$

where $\kappa = g_\mathrm{Rw} I_\mathrm{p0}$. For simplicity, we assume that $\alpha_\mathrm{s} \approx \alpha_\mathrm{p}$. The solutions to the inhomogeneous differential equation are straightforward to identify with the boundary conditions, $n_\mathrm{sf}(0) = 0$ and $n_\mathrm{sb}(L) = 0$, and are:

$$n_\mathrm{sf}(L) = \kappa L e^{-\bar{L}},$$

$$n_\mathrm{sb}(0) = \frac{\kappa}{2\alpha_\mathrm{s}}\left(1 - e^{-2\bar{L}}\right).$$

Where $\bar{L} = \alpha_s L$. Note that the photon numbers derived do not include the outcoupling efficiency, $\eta_\mathrm{out}$, in the Raman wavelength range. Thus, let the number of photons collected from the input coupler, $n_\mathrm{in} = \eta_\mathrm{out}(\lambda_\mathrm{s}) n_\mathrm{sb}(0)$, and from the output coupler, $n_\mathrm{out} = \eta_\mathrm{out}(\lambda_\mathrm{s}) n_\mathrm{sf}(L)$. Since we are taking ratios of outcoupled signals here, $\eta_\mathrm{out}(\lambda_\mathrm{s})$, does not appear in the final expressions.

Note that the optimum output from the waveguide is where $n_\mathrm{sf}(L) + n_\mathrm{sb}(0)$ is a maximum, which occurs for $L = 2L_\mathrm{sp}$.

**Indirect optical pumping by Gaussian beam**

Raman scattering into the waveguide mode can also be excited by illuminating the centre of the waveguide. In this case, part of the waveguide is illuminated with a Gaussian beam of area $A_0 = \pi \rho^2$. Here, $\rho$ is the Gaussian beam waist radius at the focal point on the sample and is defined as the $e^{-2}$ radius of a Gaussian beam when fit to the diffraction limited Airy disk function, such that

$$\rho = 0.41 \frac{\lambda}{NA}.$$

where $NA = 1.45$. The 1D optical pump profile in this case is,

$$I_\text{p}(z) = HI_{\text{g}0}\text{e}^{-2(z-L/2)^2/\rho^2}.$$

The parameter $H$ accounts for local field enhancement of the Gaussian beam due to the waveguide's nanostructured geometry. The differential equations for the Stoke's Raman scattered photons now become:

$$\frac{dn_\text{f}^{(c)}(z)}{dz} + \alpha_s n_\text{f}^{(c)}(z) = \gamma H \text{e}^{-2(z-L/2)^2/\rho^2},$$

$$\frac{dn_\text{b}^{(c)}(z)}{dz} - \alpha_s n_\text{b}^{(c)}(z) = -\gamma H \text{e}^{-2(z-L/2)^2/\rho^2},$$

where $\gamma = g_{\text{Rw}} I_{\text{g}0}$, measures the strength of Raman scattering into the waveguide mode via indirect illumination. Solving these equations requires the boundary conditions, $n_\text{sb}^{(c)}(0) = n_\text{sf}^{(c)}(L)$ (by symmetry when centrally illuminated) and $n_\text{sb}^{(c)}(L) = n_\text{sf}^{(c)}(0) = 0$. Solutions are,

$$n_\text{sf}^{(c)}(z) = \gamma H \text{e}^{-\alpha_s z} \int_0^z \text{e}^{-2(t-L/2)^2/\rho^2 + \alpha_s t} dt,$$

$$n_\text{sb}^{(c)}(z) = \gamma H \text{e}^{\alpha_s z} \left[ \int_0^L \text{e}^{-2(t-L/2)^2/\rho^2 - \alpha_s t} dt - \int_0^z \text{e}^{-2(t-L/2)^2/\rho^2 - \alpha_s t} dt \right].$$

Sub $t' = t - L/2$... and, for $\rho \ll L$, we can approximate to the standard integral with infinite limits $L \mapsto \infty$.

$$\int_0^L \text{e}^{-2(t-L/2)^2/\rho^2 \pm \alpha_s t} dz = \int_{-L/2}^{L/2} \text{e}^{-2t'^2/\rho^2 \pm \alpha_s(t'+L/2)} dz' \approx \sqrt{\frac{\pi}{2}} \rho \text{e}^{\alpha_s^2 \rho^2/8 \pm \bar{L}/2}.$$

Thus, we find our solutions for the signals at the waveguide ends under central Gaussian beam illumination.

$$n_\text{sf}^{(c)}(L) + n_\text{sb}^{(c)}(0) = \sqrt{2\pi} \rho \gamma H \text{e}^{\alpha_s^2 \rho^2/8 - \bar{L}/2}.$$

The total number of Raman photons collected from both antennas in this case is, $n^{(c)} = \eta_\text{out}(\lambda_s) \left( n_\text{sf}^{(c)}(L) + n_\text{sb}^{(c)}(0) \right).$

## 5. Local Field Enhancement by Gap Waveguide

When illuminating the waveguide centrally with a Gaussian beam, the gap structure enhances the local electric field intensity. This is considered in the above theory using the parameter $H$. Figure S4 shows a simulation of the cross section of the waveguide illuminated by a Gaussian beam. Line scans for varying distance from the gold surface were taken so that the surface enhancement could be extrapolated near the gold surface. An average surface intensity enhancement of $H = 4.29$ was determined by averaging over a line scan within the waveguide's gap region. This was normalised against the peak field of the incident Gaussian beam without the waveguide structure.

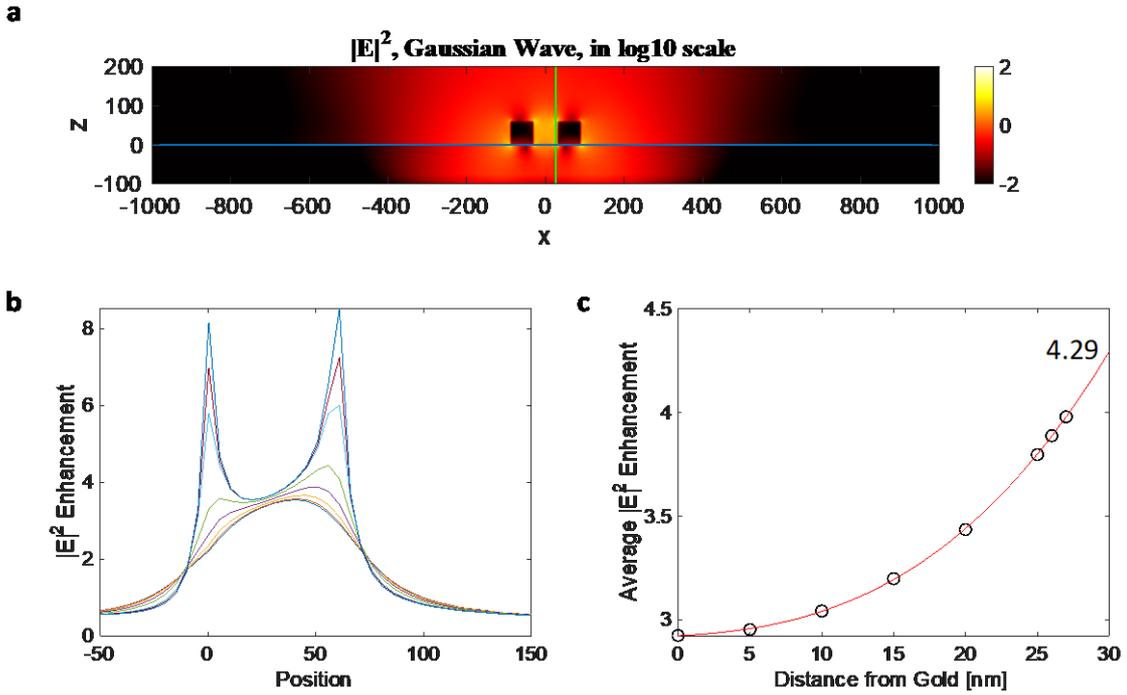

**Figure S4.** Simulation of the local electric field enhancement of a Gaussian beam polarised along the y-axis by the slot waveguide. (a) Electric field intensity in log10 scale through a cross section of Gaussian illumination beam. (b) Linescans for the positions along the x direction shown in (c). (c) Averaged intensity enhancement for each linescan within the gap region. An extrapolated average surface enhanced intensity of 4.29 is found.

## 6. Ratios of Stokes Scattering from Waveguide Outputs

We can now directly compare the Raman scattering from the ends of the waveguide under the two illumination cases. For direct illumination via the waveguide mode, the ratio of Raman scattered photons from the antennas is

$$\frac{n_{\text{out}}}{n_{\text{in}}} = \bar{L}[\sinh \bar{L}]^{-1}.$$

This equation is used in a direct comparison with experiments in the main text. The only free parameter is the attenuation coefficient of the waveguide, $\alpha_s = L_{\text{sp}}^{-1}$.

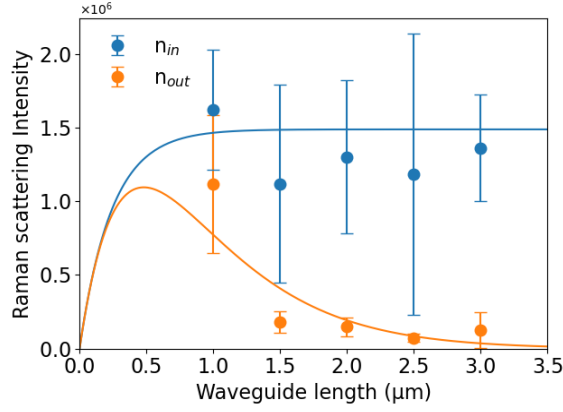

**Figure S5.** comparing $n_{\text{in}}(L)$ and $n_{\text{out}}(L)$. (The noisy ones that we did not want to put in the main manuscript.)

When comparing Raman scattered photons by the direct and indirect illumination methods, we must account for the same power, $P_0$, of illumination in both cases. For the waveguide mode, the coupled peak intensity in the waveguide is $I_{\text{p0}} = \eta_{\text{in}}(\lambda_{\text{p}}) n_{\text{g}}(\lambda_{\text{p}}) P_0 / A_{\text{wg}}(\lambda_{\text{p}})$ where $A_{\text{wg}}$ is the mode area and $n_{\text{g}}(\lambda_{\text{p}})$ is the group index at the pump wavelength. For the Gaussian beam, the peak intensity is $I_{\text{g0}} = 2P_0/A_0$, where $A_0 = \pi\rho^2$. We thus find the ratio:

$$\left(\frac{n^{(c)}}{n_{\text{in}}}\right) = \left(\frac{A_{\text{wg}}(\lambda_{\text{p}})}{A_0}\right) \frac{4\sqrt{2\pi}\alpha_s \rho H e^{\alpha_s^2 \rho^2/8 - \bar{L}/2}}{\eta_{\text{in}}(\lambda_{\text{p}}) n_g(\lambda_{\text{p}})(1 - e^{-2\bar{L}})} \approx \left(\frac{A_{\text{wg}}(\lambda_{\text{p}})}{A_0}\right) \frac{4\sqrt{2\pi}\alpha_s \rho H}{\eta_{\text{in}}(\lambda_{\text{p}}) n_g(\lambda_{\text{p}})} e^{-\bar{L}/2},$$

where the approximation holds for $L \gg \alpha_s^{-1}$ and $\rho \ll \alpha_s^{-1}$. We thus find an expression for the ratio of the Gaussian beam and waveguide mode areas:

$$\left(\frac{A_0}{A_{\text{wg}}(\lambda_{\text{p}})}\right) = \left(\frac{n_{\text{in}}}{n^{(c)}}\right) \frac{4\sqrt{2\pi}\alpha_s \rho H}{\eta_{\text{in}}(\lambda_{\text{p}}) n_g(\lambda_{\text{p}})} e^{-\bar{L}/2}.$$

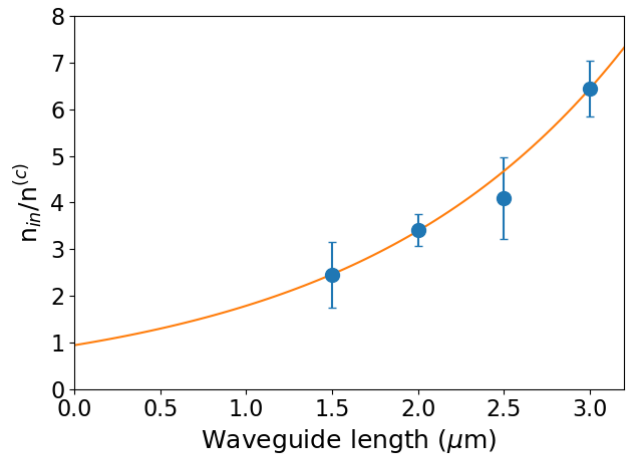

**Figure S6.** The ratio of antenna and waveguide coupled Raman signals $n_{in}/n^{(c)}$, as expressed in the equation above.

Figure S6 fits the experimental data for the ratio of Raman scattering for the two illumination cases, which is then fit to the equation:

$$\left(\frac{n_{in}}{n^{(c)}}\right) = A e^{\bar{L}/2}$$

We find A $=0.967 \pm 0.009$, and $L_{sp} = 0.780 \pm 0.004\ \mu m$.

## 7. Raman Purcell and $\beta$ Factors

We can also compute the waveguide mode's contribution to the Purcell factor at the Stoke's wavelength for the single Raman tensor component sampled (pump and emission polarised perpendicular to the gap) using the formula[1–3],

$$F_{pw} = \frac{g_{RW}}{g_R} \approx 3 \frac{n_g(\lambda_s)}{\pi} \frac{(\lambda/2)^2}{A_{wg}(\lambda_s)},$$

where the value for the group index, $n_g(\lambda_s)$, and mode area, $A_{wg}(\lambda_s)$, must be computed at the Stoke's wavelength. The total Purcell factor, $F_p$ is often approximated by assuming that the coupling to radiation is not modified, so that $F_p = 1 + F_{pw}$.

It is difficult to compute the $\beta$ factor for any spontaneous radiation process as by definition we must account for all possible channels of radiation and non-radiation. Consider the Raman scattering rate without electromagnetic enhancement, $\Gamma_{R0} \propto g_{R0}$. This rate is a sum over all possible radiation modes, $\Gamma_{R0} = \sum_i \Gamma_{R0i}$. In the waveguide, we assume that the individual radiation mode coupling strengths are not modified and that the waveguide mode coupling strength, $\Gamma_{Rw} \propto g_{RW}$, is the only additional modification to the total rate, such that,

$$\Gamma_R = \Gamma_{Rw} + \sum_i \Gamma_{R0i}.$$

We can thus estimate the Raman Purcell factor,

$$F_p \approx \frac{\Gamma_R}{\Gamma_{R0}} = \frac{\Gamma_{Rw}}{\Gamma_{R0}} + 1.$$

The fraction $\beta$ of Raman scattering coupling to a given mode is the ratio of the mode's coupling strength and the total coupling strength. The Raman $\beta$ factor for the waveguide mode here is thus $\beta = \Gamma_{Rw}/\Gamma_R$. Therefore,

$$\beta = \frac{\Gamma_{Rw}}{\Gamma_R} = \frac{\Gamma_{Rw}/\Gamma_{R0}}{\Gamma_R/\Gamma_{R0}} \approx \frac{F_{pw}}{F_{pw} + 1} = \frac{F_p - 1}{F_p}$$

This formula is often quoted in the literature as a simple estimator for the $\beta$ factor in the case of spontaneous fluorescence.

## 8. Raman Enhancement Factor

Here we compute the Raman enhancement factor of the waveguide SERS system relative to the flat gold interface control sample, as considered experimentally. The peak intensity at the gold film illuminated by a Gaussian beam of radius, $\rho_f = 0.41 \lambda_p/NA$, where $NA = 0.95$, is $I_{f0}$. The pump beam intensity on the gold film is influenced by the film's reflectivity, lowering the peak intensity at the interface by a factor $|1 - r|^2 \approx 0.16$, where $|r|^2$ is the gold film's reflectivity. Thus, $I_{f0} = 2|1 - r|^2 P_0/A_{f0}$, where $A_{f0} = \pi \rho_f^2$ is the Gaussian pump beam area on the gold reference film and $P_0$ is the incident power. Meanwhile $I_{p0} = \eta_{in}(\lambda_p) n_g(\lambda_p) P_0 / A_{wg}(\lambda_p)$, as given above. Therefore,

$$\frac{I_{p0}}{I_{f0}} = \frac{\eta_{in}(\lambda_p) n_g(\lambda_p)}{2|1-r|^2} \frac{A_{f0}}{A_{wg}(\lambda_p)}.$$

We can determine the total Raman enhancement factor by realising that $F_{pw} = |E_{wg}(\lambda_s)|^2 / |E_0(\lambda_s)|^2$ and $I_{p0}/I_{f0} = |E_{wg}(\lambda_p)|^2 / |E_0(\lambda_p)|^2$, where $|E_{wg}(\lambda_s)|^2$ and $|E_0(\lambda)|^2$ are the local electric field intensities at the pump and Stoke's wavelengths. Therefore, the Raman enhancement factor is,

$$F_R = F_{pw} \frac{I_{p0}}{I_{f0}} = \frac{|E_{wg}(\lambda_s)|^2}{|E_0(\lambda_s)|^2} \frac{|E_{wg}(\lambda_p)|^2}{|E_0(\lambda_p)|^2} R_{mol}.$$

Where $R_{mol} \approx 0.36/0.17 = 2.12$ is the ratio of the active areas of molecules illuminated in each case. Written out in full, we find,

$$F_R = \frac{3}{2\pi} \frac{\eta_{in}(\lambda_p)}{|1-r|^2} n_g(\lambda_p) n_g(\lambda_s) \left(\frac{A_{f0}}{A_{wg}(\lambda_p)}\right) \left(\frac{(\lambda/2)^2}{A_{wg}(\lambda_s)}\right) R_{mol}.$$

Note that this expression functionally resembles that given in Ref[4] from the main text. In the case of reducing the gap width further, we can now also speculate about the maximum Raman enhancement achievable in this W-SERS system. Figure S7 shows the enhancement factor for this waveguide system as a function of gap width in the range [60,10] nm, which could all be realised in practice. To estimate $R_{mol}$ as a function of gap width, we assumed a proportional reduction in theoretical propagation lengths due to the 4-ATP as for the 60 nm case considered experimentally. This is likely to be an over-estimate here, as 4-ATP dominates the absorption.

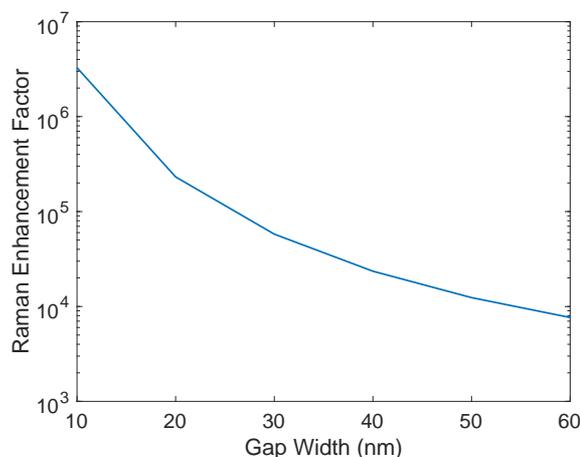

**Figure S7.** Calculated Raman Enhancement Factor as a function of the Gap Width. As in the main text, the enhancement factor is computed relative to a flat gold film functionalised with 4-ATP.

## 9. Waveguide Mode Calculations

The effective mode area and group index are required to compute waveguide SERS enhancement parameters. These were numerically calculated using Ansys Lumerical finite difference eigenmode (FDE) solver. In the simulations, the cross section of the waveguide was defined by metallic squares of side length 60 nm, positioned at the interface of a silica substrate. To avoid numerical errors due to sharp corners, the corners of the squares were rounded with radius of 5 nm. To avoid the effect of the boundary conditions on the calculated modes, the solver region was set as 600 nm in the x direction and 500 in the y direction. In the solver region, a uniform mesh region with 1 nm scale was applied. Convergence testing was used to verify all simulation settings.

The FDE solver returns the spatial mode profiles and mode properties (effective index and group index) for the fundamental gap plasmon waveguide mode. To calculate the mode area, we used the following definition,

$$A_{mode}(\boldsymbol{r}_0) = \frac{\iint W(r).dA}{1/2\varepsilon_0 |\boldsymbol{d}_0.\boldsymbol{E}(\boldsymbol{r}_0)|^2}$$

where $\boldsymbol{d}_0$ is the dipole orientation and $\boldsymbol{r}_0$ is the position. The energy density must be adjusted for the dispersive energy density in the metal. It is thus easier to substitute for the energy velocity,

$$v_g = \frac{1/2 \iint \Re\{\boldsymbol{E} \times \boldsymbol{H}^*\}.zdA}{\iint W(r).dA},$$

And compute the waveguide power flow instead. For this we also use the computed group index, $n_g = c/v_g$. To compute the waveguide mode area for the Raman scattering in this work, we averaged $A_{mode}(\boldsymbol{l})$ over the line of positions 1 nm from the metal surface along one side of the gap region where molecules were functionalised.

$$A_{wg} = \frac{1}{l_{gap}} \int_l A_{mode}(\boldsymbol{l}) dl$$

where $l_{gap} = 60\ nm$ is the length of the line element, $\boldsymbol{l}$. In all simulations the permittivities of gold and silica were adapted from the data of Johnson & Christy[5] and Palik[6].

# 10. Waveguide SERS Enhancement Parameters

Table S1 shows Raman enhancement characteristics were extracted from experimental and computational data. This data was computed from the set of parameters shown in Table S2.

**Table S1** Calculated waveguide SERS parameters determined experiment and computation.

| Parameter | Experimental | Numerical Value |
|---|---|---|
| $\dfrac{A_0}{A_{\text{wg}}(\lambda_p)}$ | $14.94 \pm 0.96$ | 14.68 |
| $\dfrac{(\lambda_p^2/4)}{A_{\text{wg}}(\lambda_p)}$ | $15.11 \pm 0.97$ | 14.85 |
| $\dfrac{I_{p0}}{I_{f0}}$ | $84.45 \pm 0.87$ | - |
| $F_p$ | - | 42.55 |
| $\beta$ | $0.993 \pm 0.001$ | 0.9765 |
| $F_R$ | $7{,}504 \pm 78$ | - |

**Table S2** Experimental and numerical parameters used in the experiment.

| Parameter | Experimental Value | Numerical Value |
|---|---|---|
| $\lambda_p$ | 790 nm | - |
| $\lambda_s$ | 800-920 nm | - |
| NA (Waveguide) | 1.45 | - |
| NA (Film) | 0.95 | - |
| $\eta_{\text{in}} = \eta_{\text{out}}$ | $29.9 \pm 1.9\%$ | 34.6% |
| $L_{\text{sp}}(\lambda_p)$ (No 4-ATP) | $2.11 \pm 0.23\ \mu m$ | $1.79\ \mu m$ |
| $L_{\text{sp}}(\lambda_p)$[1] | $0.64 \pm 0.08\ \mu m$ | - |
| $L_{\text{sp}}(\lambda_s)$ | $0.780 \pm 0.04\ \mu m$ | - |
| $\rho$ | - | $0.41\lambda/NA = 221$ nm |
| $\rho_f$ | - | $0.41\lambda/NA = 340$ nm |
| $H$ | - | 4.29 |
| $n_g(\lambda_p)$ | - | 2.64 |
| $n_g(\lambda_R)$ | - | 2.42 |
| $(\lambda_p^2/4)/A_{\text{wg}}(\lambda_p)$ | - | 14.85 |
| $(\lambda_s^2/4)/A_{\text{wg}}(\lambda_s)$ | - | 19.02 |
| $R_{\text{mol}}$ | - | 2.12 |

---

[1] The pump propagation distance was computed from the Raman measurements. The fitting to $n_{in}/n^{(c)}$ provided $L_{sp}(\lambda_s)$, while the fitting to $n_{in}/n_{out}$ provided the average $\langle L_{sp}^{-1} \rangle = \left(L_{sp}^{-1}(\lambda_s) + L_{sp}^{-1}(\lambda_p)\right)/2$, from which $L_{sp}^{-1}(\lambda_p)$ was computed.